\documentclass[%
reprint,
superscriptaddress,
amsmath,amssymb,
aps,
]{revtex4-1}

\usepackage{color}
\usepackage{graphicx}% Include figure files
\usepackage{dcolumn}% Align table columns on decimal point
\usepackage{bm}% bold math
\begin {document}
%section {title}
%\preprint{APS/123-QED}

\title{Enhancement, slow relaxation, ergodicity and rejuvenation of diffusion  in biased continuous-time random walks}% Force line breaks with \\

\author{Takuma Akimoto}
\email{takuma@rs.tus.ac.jp}
\affiliation{%
  Department of Physics, Tokyo University of Science, Noda, Chiba 278-8510, Japan
}%

\author{Andrey G. Cherstvy}
\affiliation{%
Institute for Physics \& Astronomy, University of Potsdam, 14476 Potsdam-Golm, Germany
}

%\author{Ruo Hou}
%\affiliation{%
%Institute for Physics \& Astronomy, University of Potsdam, 14476 Potsdam-Golm, Germany
%}
%\affiliation{%
%School of Mathematics and Statistics, Lanzhou University, Lanzhou 730000, China
%}

\author{Ralf Metzler}
\affiliation{%
Institute for Physics \& Astronomy, University of Potsdam, 14476 Potsdam-Golm, Germany
}


\date{\today}% It is always \today, today,

\begin{abstract}

Bias plays an important role in the enhancement of diffusion in periodic potentials. 
Using the continuous-time random walk in the presence of a bias, 
 we provide a novel mechanism for the enhancement of diffusion in a random energy landscape. 
When the variance of the waiting time diverges, in contrast to the bias-free case the dynamics with bias becomes superdiffusive. In the 
superdiffusive regime,  we find a distinct initial ensemble dependence of the diffusivity. 
 We show that the time-averaged variance converges to the corresponding ensemble-averaged variance, i.e., ergodicity is preserved. 
 However, trajectory-to-trajectory fluctuations of the time-averaged variance decay slowly. 
 Our finding suggests that in the superdiffusive regime the diffusivity for a non-equilibrium initial ensemble gradually  increases 
 to that for an equilibrium ensemble when the start of the measurement is delayed, corresponding to a rejuvenation of diffusivity. 
\end{abstract}

%\pacs{05.45.Ac, 05.40.Fb, 87.15.Vv}% PACS, the Physics and Astronomy
\maketitle

%\tableofcontents

\section{introduction}
Mixing of more than two fluids is a key operation of microfluidic devices in chemistry, biology and industry, 
in which diffusion is an essential mechanism for mixing \cite{Knight1998,Todd2005,Kim2010}.
In particular, achieving an enhancement of the diffusivity is pivotal for mixing of particles in heterogeneous environments 
because diffusion in such systems is often slow.  
One of the most applicable controls of the diffusivity is adding a directed external force, i.e., a bias. In discrete-time random walks
the bias, characterized  by the difference between the probabilities of right and left jumps, suppresses 
the diffusivity.  In particular, the variance of the displacement grows linearly with time \cite{Feller1968}:
 ${\rm Var}(x_n) \equiv \langle x_n^2 \rangle - \langle x_n \rangle^2=4pqn$,
where $p$ and $q=1-p$ are the probabilities of right and left jumps, respectively, and we assume the jump size 
is fixed to unity. 
Thus, the diffusivity defined by $D\equiv {\rm Var}(x_n)/n$ is given by 
$D= (1-\varepsilon)(1+\varepsilon)$,
where $\varepsilon = p-q$. 
In the absence of a bias ($\varepsilon=0$) $D$ is maximized. In other words, the diffusivity is suppressed 
by the bias for discrete-time random walks. 

This trend may in fact be reversed when the time steps are continuous random variables. 
It is well-known that the diffusivity can be enhanced by an external field in diffusion in periodic potentials, i.e., 
tilted sinusoidal potentials \cite{Reimann2001,Reimann2002}. In particular, when the diffusivity in absence of an external force 
is small due to deep periodic potential wells or low temperatures, 
the diffusivity characterized by the variance of the displacement is greatly accelerated
(``giant acceleration") at an optimal external force. As seen above, the external force actually suppresses the diffusivity 
due to a directed motion  in discrete-time random walks. Conversely, the bias decreases the escape time from a valley of a periodic 
potential, which contributes to the enhancement of diffusivity. With the aid of this  trade-off relation, 
the diffusivity can be maximized at some optimal external force. As the diffusivity enhancement  by a bias
 is a universal phenomenon in periodic potentials, many experiments have been designed to 
  realize this enhancement \cite{Mykhaylo2008,Reimann2008,Hayashi2015,Ma2015,Kim2017,Ma2017}. 
 
 Effects of a bias in many-body systems has also attracted a considerable interest aiming 
 to unravel non-equilibrium properties \cite{Habbdas2004, Whilhelm2008, Gazuz2009, Shin2015}. 
In particular, it is essential to investigate the effect of an external force on the diffusion of a particle 
in many-particle systems \cite{Harrer2012, Gruber2016}.
 In many crowding systems, such as diffusion in cells and active diffusion of colloidal particles, diffusion becomes anomalous, i.e., 
 the mean square displacement  does not increase linearly with time \cite{Richert-2002,Golding2006,Weigel2011,Hofling2013}. 
 Recently, it became known that an external field in crowding systems induces superdiffusion, i.e., $\langle \bm{r} (t)^2 \rangle 
 \propto t^{\beta}$ with $\beta >1$ \cite{Schroer2013,Benichou2013, reverey2015, Gradenigo2016,Leitmann2017}. As a mean-field approach, 
 the continuous-time random walk (CTRW) is often used for quenched heterogeneous environments. In fact, 
 the field-induced superdiffusion may be observed when a bias is added in the CTRW \cite{Burioni2013,Burioni2014}. 
 This field-induced phenomenon is essential to unravel the enhancement of the diffusivity in heterogeneous systems. 
 
 In this Rapid Communication, we investigate an initial-ensemble dependence of the variance of the displacement and 
 ergodic properties of the time-averaged variance of the displacement in the CTRW with drift: 
 the variance becomes superdiffusive  when the variance of the waiting time diverges. Therefore, it is a fundamental question 
 whether the system is still ergodic. If the system is ergodic, it is important to clarify how the initial ensemble difference  affects
 the relaxation process.

\section{model}
The quenched trap model (QTM) is used to describe a random walk in a quenched 
random potential landscape, in which the depths of the potential wells are randomly 
distributed \cite{bouchaud90}.  When the distribution of depths follows an exponential law, the waiting-time distribution 
is of power-law form \cite{Bardou2002}, where the power-law exponent depends on the temperature. 
CTRW is a random walk with random  waiting times, corresponding to 
an annealed model of the QTM. In CTRW, waiting times are independent and identically distributed random variables, which do not 
depend on the site. On the other hand, waiting-time distributions clearly depend on the site in the QTM. In this sense, the
CTRW is homogeneous and sometimes fails to capture the rich physical properties due to the quenched disorder \cite{Bertin2003, Burov2007, Miyaguchi2011, *Miyaguchi2015, Luo2015, Akimoto2016}. 
However, the CTRW is a good approximation when the spatial dimension is equal and greater than two \cite{Machta1985} 
or in the presence of a bias minimizing the risk of back stepping \cite{Scher1975}. 

To investigate the effects of the bias on the diffusive properties in heterogeneous environments, 
we consider a CTRW with a drift. We assume that the waiting-time distribution follows a power-law distribution
\begin{equation}
 \psi(\tau) \sim \alpha \tau_0^{\alpha} \tau^{-1-\alpha} \quad (\tau \gg \tau_0).
\end{equation}
The Laplace transform for the case $\alpha >1$ considered here reads
\begin{equation}
\hat{\psi}(s) =
\left\{
\begin{array}{ll}
%1 -cs^\alpha +o(s^\alpha) &(\alpha<1)\\
1-\mu s  +cs^\alpha +o(s^\alpha) &(1<\alpha<2)\\
\\
1-\mu s  +\frac{1}{2}(\sigma^2+\mu^2) s^2 +o(s^2) &(2<\alpha)\\
\end{array}
\right.,
\end{equation}
where  the mean and the variance of the waiting time are denoted by $\mu$ and $\sigma^2$, respectively, and $c=|\Gamma(1-\alpha)|\tau_0^\alpha$.  
For $\psi(\tau) = \alpha \tau_0^{\alpha} \tau^{-1-\alpha}$ \quad ($\tau \geq \tau_0$), $\mu$ and $\sigma^2 + \mu^2$ are 
given by $\alpha \tau_0 /(\alpha-1)$ and $\alpha \tau_0^2 /(\alpha -2)$, respectively.

Let $N_t$ be the number of jumps of a random walker until 
time $t$. Then, we have the first moment of displacement $x(t)$ with $x(0)=0$ as
\begin{equation}
\langle x(t) \rangle =(p-q) \langle N_t \rangle.
\label{mean_displacement}
\end{equation}
%where $p$ is probability for the right jump and $q=1-p$. 
The variance of the displacement,  ${\rm Var}(x(t))\equiv \langle x(t)^2 \rangle - \langle x(t) \rangle^2$, is expressed
through $N_t$:
\begin{equation}
{\rm Var}(x(t)) %\equiv \langle x(t)^2 \rangle - \langle x(t) \rangle^2 
=(p-q)^2(\langle N_t^2 \rangle - \langle N_t \rangle^2) 
+4pq \langle N_t \rangle.
\label{var_displacement}
\end{equation}
Moreover, the variance of the displacement from $t$ to $t+\Delta$, i.e., $\delta x(t,t+\Delta) \equiv x(t+\Delta) -x(t)$, is given by
\begin{align}
&{\rm Var}(\delta x (t,t+\Delta)) \nonumber\\%&\equiv \langle \delta x (t,t+\Delta)^2 \rangle - \langle \delta x (t,t+\Delta) \rangle^2 \\
%&= {\rm Var}(x_{t+\Delta}) + {\rm Var}(x_t) - 2 (\langle x_{t+\Delta} x_t \rangle -\langle x_{t+\Delta} \rangle \langle x_t \rangle )\\
%&= \langle x_{t+\Delta}^2 \rangle -\langle x_t^2 \rangle - 2 \langle \delta x (t,t+\Delta) x_t \rangle -\langle x_{t+\Delta} \rangle^2
%+ \langle x_{t} \rangle^2 +2\langle \delta x (t,t+\Delta) \rangle \langle x_t \rangle \\
&=(p-q)^2(\langle N_{t+\Delta}^2 \rangle - \langle N_t^2 \rangle  -  2 \langle N_{t,t+\Delta} N_t \rangle) \nonumber\\
& - (p-q)^2 (\langle N_{t+\Delta} \rangle^2 - \langle N_t \rangle^2
  - 2\langle N_{t,t+\Delta} \rangle \langle N_t \rangle)\nonumber\\
&  +4pq (\langle N_{t+\Delta} \rangle -\langle N_t \rangle) , 
\label{var_t}
 \end{align}
 where $N_{t,t+\Delta}=N_{t+\Delta}-N_t$ is the number of jumps in $[t,t+\Delta]$. Therefore, 
 the mean and variance of the displacement can be calculated using the moments and the correlation of $N_t$ 
  obtained from renewal theory \cite{cox}.
 
 Here, we consider two typical renewal processes, i.e., ordinary and equilibrium renewal processes \cite{cox}. 
 Renewal processes are point processes in which the time intervals between successive renewals are independent and 
 identically distributed random variables. In CTRW, the distribution corresponds to the waiting-time distribution $\psi(\tau)$.
 
 One has to be careful on the first renewal event because the time when 
 the first renewal occurs is a random variable, but the distribution is not the same as $\psi(\tau)$, in general \cite{cox,God2001,Schulz2013,*Schulz2014}. 
 An ordinary renewal process is a renewal process in which the distribution of the time when 
 the first renewal occurs follows $\psi(\tau)$ \cite{cox}. In other words, a renewal occurs 
 at the time when the observation starts. In equilibrium renewal processes, a measurement starts after the system has evolved for a long time, 
 and thus the distribution of 
 the first renewal time is not the same as $\psi(\tau)$, except for the case when the distribution 
 follows an exponential law. When the mean waiting time exists ($\mu < \infty$), the first renewal time distribution can be represented by 
 \cite{cox,God2001}
 \begin{equation}
 \psi_0(\tau) = \mu^{-1} \int_\tau^\infty \psi(\tau')d\tau'.
 \label{fwt_eq}
 \end{equation}

% In equilibrium process ($\alpha>1$),
%\begin{align}
%{\rm Var}(\delta x (t,t+\Delta)) &=(p-q)^2(\langle N_{t+\Delta}^2 \rangle_e - \langle N_t^2 \rangle_e  -  2 \langle N_{t,t+\Delta} N_t \rangle_e) 
% +4pq \frac{\Delta}{\mu}-(p-q)^2 \frac{\Delta^2}{\mu^2}.
%\end{align}

\section{Variance of the displacement}
%\subsection{the case the variance exists ($\alpha >2$)}
The first moment $\langle N_t \rangle$, called the renewal function, is well-known in renewal theory \cite{cox, God2001}:
for $\alpha>1$ it becomes $\langle N_t \rangle \sim t/\mu$ for $t \gg \tau_0$. In particular, it is exact,  $\langle N_t \rangle = t/\mu$ 
for $t>0$ when the first renewal time follows the equilibrium distribution~(\ref{fwt_eq}). 
Using Eq.~(\ref{mean_displacement}) and the renewal function, we have the mean displacements
\begin{equation}
\langle x(t) - x(0) \rangle_{\rm eq} = (\varepsilon /\mu) t
\end{equation}
for $t>0$ and 
\begin{equation}
\langle x(t) - x(0) \rangle_{\rm or} \sim (\varepsilon /\mu) t
\end{equation}
 for $t\gg \tau_0$  in equilibrium ($\langle \cdot \rangle_{\rm eq}$) and ordinary ($\langle \cdot \rangle_{\rm or}$) 
 renewal processes, respectively.
  %Here,  we denote the equilibrium and ordinary renewal processes as 
 %$\langle \cdot \rangle_{\rm eq}$ and $\langle \cdot \rangle_{\rm or}$, respectively, when we specify the initial ensemble.
 
The second moment of $N_t$ is also well known in renewal theory \cite{cox}. For $\alpha >2$, the asymptotic behavior 
of the variance of $N_t$ is not affected by the initial ensemble and is given by 
\begin{equation}
{\rm Var} (N_t) \equiv \langle N_t^2 \rangle  -\langle N_t \rangle^2 =  \frac{\sigma^2}{\mu^3}t + o(t). 
\end{equation}
Therefore, the variance of the displacement for both ordinary and equilibrium processes becomes 
\begin{equation}
{\rm Var}(x(t)) \sim \left(\frac{1}{\mu} + \frac{\sigma^2 -\mu^2}{\mu^3}\varepsilon^2\right)t 
\end{equation}
 for $t\gg \tau_0$ and $\alpha > 2$.  Interestingly,  the diffusivity 
\begin{equation}
D_\varepsilon \equiv \frac{1}{\mu} + \frac{\sigma^2 -\mu^2}{\mu^3}\varepsilon^2
\end{equation}
 is enhanced by the bias $\varepsilon$ 
when $\sigma^2 > \mu^2$. The critical value $\alpha_c$, i.e., $\sigma^2 = \mu^2$ at $\alpha=\alpha_c$, 
is given by $\alpha_c=1+\sqrt{2}$. Therefore, the diffusion is enhanced when 
$2<\alpha<\alpha_c$ (Fig.~1): 
the variance $\sigma^2$ of the waiting time is greater than $\mu^2$ and an enhancement of the diffusivity 
due to the external force is achieved. To the best of our knowledge, this enhancement mechanism, which is completely different from diffusion 
in tilted periodic potentials \cite{Reimann2001,Reimann2002}, has not been clarified so far. Therefore, 
this new mechanism of the diffusion enhancement is one of our main results. 
%For $\alpha<3$, the order of the subleading term becomes 
%$O(t^{3-\alpha})$. More precisely, the subleading term is given by $\frac{2c}{\mu^3 \Gamma(4-\alpha)}t^{3-\alpha}$.
 %Therefore, it cannot be neglected, especially for $\alpha \cong 2$.

To obtain ${\rm Var}(\delta x (t,t+\Delta))$, we need to calculate the correlation $\langle N_t N_{t,t+\Delta}  \rangle$. 
By a similar calculation as in Ref.~\cite{Barkai2007}, one can obtain the Laplace transform of $\langle N_t N_{t,t+\Delta}\rangle$ with 
respect to $t$ and $\Delta$ as
\begin{equation}
\underset{u,s}{\mathcal{L}} \langle N_tN_{t,t+\Delta} \rangle_{\rm eq} = \frac{\hat{\psi}(s) - \hat{\psi}(u)}{\mu us (u-s)[1-\hat{\psi}(u)][1-\hat{\psi}(s)]}
\end{equation}
and
\begin{equation}
\underset{u,s}{\mathcal{L}} \langle N_tN_{t,t+\Delta} \rangle_{\rm or} = \frac{[\hat{\psi}(s) - \hat{\psi}(u)]\hat{\psi}(u)}{s (u-s)[1-\hat{\psi}(u)]^2[1-\hat{\psi}(s)]}.
\end{equation}
%for equilibrium and ordinary renewal processes, respectively. 
By the inverse Laplace  transform, we have 
\begin{equation}
\langle N_tN_{t,t+\Delta}\rangle \sim \frac{t\Delta}{\mu^2} +  \frac{(\sigma^2 - \mu^2) \Delta}{2\mu^3} 
 +o(\Delta)
\end{equation}
for $\alpha>2$ and $\tau_0 \ll \Delta \ll t$, which is valid for both types of renewal processes. Therefore, 
the variance %of $\delta x(t,t+\Delta) \equiv x(t+\Delta) -x(t)$ 
becomes
\begin{align}
{\rm Var}(\delta x (t,t+\Delta)) & \sim 
%(p-q)^2 \left(\langle N_{t+\Delta}^2 \rangle - \langle N_t^2 \rangle  -  \frac{2\Delta t}{\mu^2} -(\sigma^2 - \mu^2) \frac{\Delta}{\mu^3}\right) 
%+4pq \frac{\Delta}{\mu} -(p-q)^2 \frac{\Delta^2}{\mu^2}\\
\varepsilon^2 \frac{\sigma^2}{\mu^3}\Delta  + (1-\varepsilon^2) \frac{\Delta}{\mu}
={\rm Var}(x_\Delta)
\label{var2}
\end{align}
and is stationary, i.e., independent of $t$ in the asymptotic limit. 
%We note that the Einstein relation, i.e., $\langle x(t) \rangle_{\varepsilon\ne 0} = \varepsilon 
%\langle x(t)^2 \rangle_{\varepsilon=0}$, is valid.

%%%%%%%%%%%%%%%%%%%%%%%%%%%%%%%%%%%%%%%%%%%%%%%%%%%%%%
\begin{figure}
\includegraphics[width=.9\linewidth, angle=0]{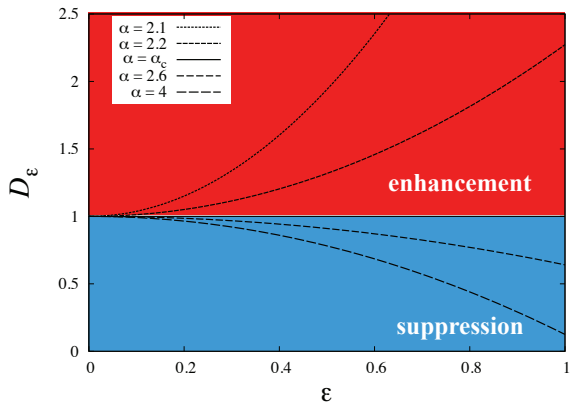}
\caption{Effect of bias $\varepsilon$ on diffusivity for different $\alpha$, where the mean is set to unity ($\mu=1$). Solid and dashed curves 
are the theoretical result, Eq.~(\ref{var2}), i.e., $D=1+(\sigma^2 -1)\varepsilon^2$. Diffusivity can be enhanced by the bias 
for $\alpha > \alpha_c$.}
\label{eav_rejuvenation}
\end{figure}
%%%%%%%%%%%%%%%%%%%%%%%%%%%%%%%%%%%%%%%%%%%%%%%%%%%%%%

%\subsection{the case the mean exists ($1<\alpha <2$)}
Unlike the asymptotic behavior of the variance for $\alpha > 2$, it depends on the initial ensemble
for $1<\alpha <2$. This is due to the fact that the variance of $N_t$ is 
\begin{equation}
\langle N_t^2 \rangle_{\rm or} -\langle N_t \rangle_{\rm or}^2 =  (\alpha-1) D(\alpha) t^{3-\alpha} + o(t^{3-\alpha}),
\end{equation}
and
\begin{equation}
\langle N_t^2 \rangle_{\rm eq} -\langle N_t \rangle_{\rm eq}^2 = D(\alpha) t^{3-\alpha} + o(t^{3-\alpha}),
\end{equation}
%for ordinary and equilibrium renewal processes, respectively, 
where $D(\alpha)=2c\mu^{-3}/\Gamma(4-\alpha)$.
It follows that the variances of the displacement in ordinary and equilibrium renewal processes are given by
\begin{equation}
{\rm Var}(x (t))_{\rm or} = \varepsilon^2 (\alpha-1) D(\alpha) t^{3-\alpha} + 4pq \frac{t}{\mu} + o(t),
\end{equation}
and
\begin{eqnarray}
{\rm Var}(x (t))_{\rm eq} = \varepsilon^2  D(\alpha) t^{3-\alpha} 
+ 4pq\frac{t}{\mu} + o(t),
\label{eav_eq}
\end{eqnarray}
respectively. Therefore, the spreading of particles with respect to the mean becomes superdiffusive with exponent ($3-\alpha$). 
However, it should be noted that the coefficients of the leading terms differ by a factor $(\alpha-1)$ according to 
 the initial ensemble. 
This initial-ensemble dependence is sometimes observed for the case when the second moment of the waiting time 
diverges  \cite{Akimoto2007,Akimoto2012,Godec2013, *Godec2013a, Froemberg2013}.  

For $1<\alpha < 2$ and $\Delta \ll t$, the correlation $\langle N_t N(t,t +\Delta) \rangle$ is given by
\begin{eqnarray}
\langle N_t N_{t,t+\Delta} \rangle_{\rm or} &\sim& \frac{t\Delta}{\mu^2} + \frac{2ct^{2-\alpha} \Delta}{\mu^3 \Gamma(3-\alpha)}
-  \frac{c \Delta^{3-\alpha}}{\mu^3 \Gamma(4-\alpha)} \nonumber\\
&+& \frac{ct^{1-\alpha}\Delta^2}{2\mu^3 \Gamma(2-\alpha)}+o(t^{1-\alpha}\Delta^2).
\end{eqnarray}
It follows that for the ordinary renewal process the variance of $\delta x(t,t+\Delta)$ with $t \gg \Delta$ becomes 
\begin{align}
{\rm Var}(\delta x (t,t+\Delta))_{\rm or} 
&\sim \varepsilon^2 D(\alpha) \Delta^{3-\alpha} - \frac{\varepsilon^2 c \Delta^2}{\mu^3 \Gamma(2-\alpha)t^{\alpha-1}}
\nonumber\\
&+ 4pq \left(\frac{\Delta}{\mu} 
+ \frac{(2-\alpha)c\Delta }{\mu^2 \Gamma (3-\alpha) t^{\alpha -1}} \right) .
%\to {\rm Var}(x_\Delta).
\label{var12}
\end{align} 
Therefore,   the variance of $\delta x(t,t+\Delta)$ for the ordinary renewal process has a clear $t$ dependence  and 
approaches that of the equilibrium renewal process (see Fig.~2). 
Although the dependence of the variance on the aging time $t$ gradually disappears, i.e., 
${\rm Var}(\delta x (t,t+\Delta))_{\rm or}  \to {\rm Var}(x (\Delta))_{\rm eq}$ for $t\to \infty$, the aging effect lasts for a long time 
when $\alpha$ is close to one. This recovery of diffusivity ({\it rejuvenation of diffusivity}) is enhanced by the increase 
of the aging time. In other words, if one waits to start measurements for a long time, the observed diffusivity 
approaches the diffusivity of the  equilibrium initial ensemble, which is enhanced by the factor $(\alpha -1)^{-1}$.  
 This enhancement becomes significant especially for $\alpha \to 1$. 
%Note that the Einstein relation is also valid in this case. 

%%%%%%%%%%%%%%%%%%%%%%%%%%%%%%%%%%%%%%%%%%%%%%%%%%%%%%
\begin{figure}
\includegraphics[width=.6\linewidth, angle=-90]{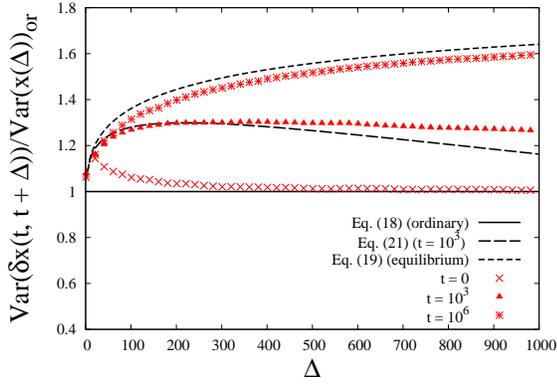}
\caption{Ensemble-averaged variance of the displacement $\delta x(t, t+\Delta)$ divided by 
${\rm Var} ( x(\Delta))_{\rm or}$ for different $t$. 
Symbols: numerical simulations, in which we generated waiting times according to the PDF 
$\psi (\tau) = \alpha \tau^{-1-\alpha}$ ($\tau\geq1$). Solid lines: Eq.~(\ref{var12}).
The variance increases with growing $t$ 
and approaches that of the equilibrium renewal process, Eq.~(\ref{eav_eq}).}
\label{eav_rejuvenation}
\end{figure}

\section{time-averaged variance of the displacement}

Here, we define the time-averaged variance (TAV) of the displacement as 
\begin{equation}
\overline{{\rm Var}(x_\Delta; t)} = \frac{1}{t-\Delta} \int_0^{t-\Delta} \left(\delta x (t', t'+\Delta) -  \frac{\varepsilon\Delta}{\mu} \right)^2 dt',
\label{TAV_def}
\end{equation}
where we have already observed  $\varepsilon \Delta /\mu = \langle \delta x (t',t' + \Delta) \rangle_{\rm eq}$, which does not depend on $t'$. 
The variable $t$ in Eq.~(\ref{TAV_def}) now corresponds to the total measurement time, $\Delta$ is the lag time, and the overline denotes time averaging.
Expanding the integrand, the TAV can be written as 
\begin{eqnarray}
\overline{{\rm Var}(x_\Delta; t)} &-& \left(\frac{\Delta}{\mu}\right)^2 \varepsilon^2 = \frac{1}{t-\Delta} \int_0^{t-\Delta} \delta x (t', t'+\Delta)^2 dt' \nonumber\\
 &-&  \frac{2\varepsilon\Delta}{\mu(t-\Delta)} \int_0^{t-\Delta}  \delta x (t', t'+\Delta)dt'.
 \label{tav_eq}
\end{eqnarray}
%The ensemble average of the TAV is given by
%$\left\langle \overline{{\rm Var}(x_\Delta; t)} \right\rangle 
%&= \frac{1}{t-\Delta} \int_0^{t-\Delta} \left\langle \left(\delta x (t',t'+\Delta) - 
%\langle \delta x (t',t'+\Delta) \rangle\right)^2 \right\rangle dt'\\
%= \frac{1}{t-\Delta} \int_0^{t-\Delta} {\rm Var} (\delta x (t',t'+\Delta)) dt'.$
As follows from Eqs.~(\ref{var2}) and (\ref{var12}), the ensemble average of the TAV converges to a constant for $\alpha > 1$, 
\begin{equation}
\left\langle \overline{{\rm Var}(x_\Delta; t)} \right\rangle \to {\rm Var}  (x(\Delta))_{\rm eq}
\label{tav_conv}
\end{equation}
as $t\to \infty$. In particular, the ensemble average of the TAV for the ordinary renewal process becomes 
\begin{eqnarray}
\left\langle \overline{{\rm Var}(x_\Delta; t)} \right\rangle_{\rm or} &-&   {\rm Var}  (x(\Delta))_{\rm eq}
\sim  K(\alpha) t^{1-\alpha},
\end{eqnarray}
where $K(\alpha)=\frac{\varepsilon^2 c \Delta^2}{\mu^3 \Gamma(2-\alpha)}
 \left( \frac{\mu}{\varepsilon^2 \Delta} -1 \right)$ for $1<\alpha <2$. 
%In equilibrium renewal process, the TAV can be written as 
%In the large $t$ limit, this expression is valid for ordinary renewal process. In what follows, we use Eq.~(\ref{TAV_def}) as 
%the definition of the TAV. 

%On the other hand, the ensemble average of the TAV shows aging for $\alpha <1$:
\if0
\begin{equation}
\left\langle \overline{{\rm Var}(x_\Delta; t)} \right\rangle \sim 
\left\{ \frac{2\varepsilon^2  \Delta^{1+\alpha}}{c^2 \Gamma(2+\alpha) \Gamma(\alpha + 1)} + \frac{4 pq\Delta}{c\Gamma(\alpha +1)}\right\}t^{\alpha-1}.
\end{equation}
\fi

The time average of the displacement can be approximated by
$\int_0^{t-\Delta}  \delta x (t', t'+\Delta)dt'/(t-\Delta) \sim \sum_{k=1}^{N_t} z_k  \Delta/t$ 
for $t\gg \Delta$, where $z_k$ is the $k$-th jump ($z_k=\pm 1$). By the law of large numbers, i.e., 
$\sum_{k=1}^n z_k/n \to \langle z_k \rangle = \varepsilon$ for $n\to \infty$, we have
\begin{equation}
\frac{1}{t-\Delta} \int_0^{t-\Delta}  \delta x (t', t'+\Delta)dt' \sim \frac{N_t}{t} \varepsilon  \Delta.
\end{equation}
Here, we use a similar approximation for the squared displacements invented in Ref.~\cite{Miyaguchi2013} 
(see also the argument in Refs.~\cite{He2008, Schulz2013, *Schulz2014}). While this approximation 
is used for the CTRW without bias, it is also valid for the CTRW with bias \cite{He2008}.
Therefore, we have
\begin{equation}
\frac{1}{t-\Delta} \int_0^{t-\Delta}  \delta x (t', t'+\Delta)^2 dt' \sim \frac{N_t}{t} ( \Delta + \varepsilon^2 h(\Delta)),
\end{equation}
where $h(\Delta)$ is a function of $\Delta$. 
 For $\alpha>1$ ($\mu < \infty$), the TAV becomes 
\begin{align}
 \overline{{\rm Var}(x_\Delta; t)}  - \varepsilon^2 \left(\frac{\Delta}{\mu}\right)^2  \sim \frac{N_t}{t} 
 H(\Delta),
 \label{var-nt}
\end{align}
where $H(\Delta) = \Delta + \varepsilon^2 h(\Delta) - \frac{2 \varepsilon^2 \Delta^2}{\mu}$. 
Taking the ensemble average of Eq.~(\ref{var-nt}) and using Eq.~(\ref{tav_conv}) lead to 
$
h(\Delta) = \frac{\sigma^2 -\mu^2}{\mu^2} \Delta +   \frac{\Delta^2}{\mu} 
$
and
$
h(\Delta) = \mu D(\alpha) \Delta^{3-\alpha} -  \Delta +  \frac{\Delta^2}{\mu} 
$
for $\alpha > 2$ and $1<\alpha<2$, respectively. We confirmed numerically that this relation is valid only for 
$\alpha <2$. For $\alpha >2$, the ensemble average is given by
$\langle H(\Delta) \rangle = \Delta + \varepsilon^2 h(\Delta) - \frac{2 \varepsilon^2 \Delta^2}{\mu}$ but an additional term 
is needed in $H(\Delta)$ and will be considered in detail elsewhere.

To characterize the relaxation process, we consider the relative standard deviation (RSD) \cite{Uneyama2012, *Uneyama2015} of the TAV 
\begin{equation}
\Sigma(t;\Delta)\equiv \frac{\sqrt{\langle \{\overline{ {\rm Var}(x_\Delta;t)}\}^2 \rangle  - \langle \overline{{\rm Var}(x_\Delta;t)}} \rangle^2}
{\langle \overline{{\rm Var}(x_\Delta;t)} \rangle}.
\label{rsd}
\end{equation} 
This is the squared root of the ergodicity breaking parameter, which is widely used to investigate  ergodic 
properties \cite{He2008, Metzler2014}.
For $1<\alpha <2$ using Eq.~(\ref{var-nt}) we have
\begin{equation}
\Sigma(t;\Delta) \sim \sqrt{\frac{\langle N_t^2 \rangle  - \langle N_t \rangle^2}{t^2}}
\frac{| H(\Delta)|}{ {\rm Var}(x_\Delta)}.
%\label{rsd}
\end{equation} 
Therefore, the RSD for $\tau_0 \ll \Delta \ll t$ becomes 
\begin{equation}
\Sigma(t;\Delta)\sim %\left\{
%\begin{array}{ll}
%\dfrac{\sigma}{\sqrt{\mu}} \left| 1 - \dfrac{\varepsilon^2  \Delta}{\mu + (\sigma^2 - \mu^2)\varepsilon^2} \right| t^{-\frac{1}{2}} &(2<\alpha)\\
%\\
\sqrt{D(\alpha)} \left|\mu - \dfrac{\varepsilon^2 \Delta^2}{\mu {\rm Var}(x_\Delta)} \right|
t^{-\frac{\alpha-1}{2}} %&(1<\alpha <2)
%\\
%\sqrt{\dfrac{2\Gamma(1+\alpha)^2}{\Gamma(1+2\alpha)} - 1} &(\alpha <1).
%\end{array}
%\right.
\end{equation} 
for the equilibrium initial ensemble. For $1<\alpha <2$, the RSD decays as $t^{-\frac{\alpha -1}{2}}$, which is anomalously 
slower than the usual case, $t^{-\frac{1}{2}}$. 
Therefore, trajectory-to-trajectory fluctuations of the time-averaged variance remain large even for long measurement times.
This slow relaxation has been also discussed in CTRW without bias \cite{Akimoto2011}.

\section{Conclusion}

The diffusivity for diffusion processes with drift is characterized by the variance of the displacement. Using the paradigmatic CTRW model 
with drift, we uncovered
a novel mechanism of diffusivity enhancement in heterogeneous evironments, which is supported by an increased variance of waiting times. 
In particular, the diffusivity becomes infinite (superdiffusive) when the variance diverges. In the superdiffusive regime,
we found an intrinsic difference of the diffusivity due to the initial ensembles, e.g., for ordinary and equilibrium renewal processes. 
This initial-ensemble dependence is essential when the second moment of the waiting time diverges. For the ordinary renewal process, 
the diffusivity increases approaching that of the corresponding equilibrium  process according to the increase of the aging time. 
In other words, if one waits to measure the diffusivity for a long time, the observed diffusivity is greatly enhanced compared to the diffusivity 
measured immediately. 
 This recovering of diffusivity has a significant implication of rejuvenation in superdiffusive physical systems.
We also showed that TAVs converge to a constant, which is given by the ensemble-averaged variance with the equilibrium initial ensemble.
Therefore, the system is ergodic whereas there is a distinct  dependence of the ensemble-averaged variance on the initial ensemble. 
Finally, we  found that trajectory-to-trajectory 
fluctuations of the TAVs decay anomalously slow, as compared to standard random walks. 

%section {bibliography}

%\bibliography{fis}


\end{document}